\documentclass[11pt]{article}
\usepackage[colorlinks=true]{hyperref}
\hypersetup{colorlinks=true,linkcolor=teal,citecolor=blue,urlcolor=blue}
\usepackage[miktex]{gnuplottex}
\usepackage{pgfplots}
\usepackage{graphicx}
\usepackage{amsmath}
\usepackage{amssymb}
\usepackage{latexsym}
\usepackage{color}
\usepackage{subfigure}
\usepackage{pst-all}
\usepackage{pstricks,pst-node,pst-text,pst-3d}
\usepackage{ifpdf}
\usepackage{multimedia}
\usepackage{lmodern}
\usepackage{cite}
\usepackage{dsfont}
\usepackage{bbm}
\usepackage{lipsum}
\usepackage{url}
\usepackage{textcomp}
\usepackage{bbold}
\usepackage{epstopdf}
\usepackage{epsfig}
\usepackage{mathtools}
\usepackage{nccmath}
\usepackage{float}
 \usepackage[normalem]{ulem}
\usepackage{ifpdf}
\ifpdf
\else
  \usepackage{pstricks}
\fi
  \hypersetup{}
\newcommand{\bra}{\begin{array}}
\newcommand{\era}{\end{array}}
\newcommand{\beq}{\begin{equation}}
\newcommand{\eeq}{\end{equation}}
\newcommand{\beqar}{\begin{eqnarray}}
\newcommand{\eeqar}{\end{eqnarray}}

\def\BC{\bb C}
\def\_\BC{\bbi C}

\def\( {\left(}
\def\) {\right)}
\def\[ {\left[}
\def\] {\right]}
\def\no2 {{\textstyle{n\over 2}}}

\def \D {{\mathcal D}}

\begin{document}
\thispagestyle{empty}
\begin{center}

\vspace{1.8cm}
\renewcommand{\thefootnote}{\fnsymbol{footnote}}
 {\Large {\bf Tripartite quantum-memory-assisted entropic uncertainty relations for multiple measurements}}\\

\vspace{1.5cm} {\bf Hazhir Dolatkhah}$^{1,2}${\footnote {
email: {\sf h.dolatkhah@gmail.com}}}, {\bf Saeed Haddadi}$^{3,4}$, {\bf Soroush Haseli}$^{5}$, {\bf Mohammad Reza Pourkarimi}$^{6}$ and {\bf Mario Ziman}$^{1,7}$\\
\vspace{0.5cm}

$^{1}${\it RCQI, Institute of Physics, Slovak Academy of Sciences, \\D\'{u}bravsk\'{a} cesta 9, 84511 Bratislava, Slovakia}\\ [0.5em]
$^{2}${\it Department of Physics, University of Kurdistan, P.O.Box  66177-15175, Sanandaj, Iran}\\ [0.5em]
$^{3}${\it Faculty of Physics, Semnan University, P.O.Box 35195-363, Semnan, Iran}\\ [0.5em]
$^{4}${\it Saeed's Quantum Information Group, P.O.Box 19395-0560, Tehran, Iran}\\[0.5em]
$^{5}${\it Faculty of Physics, Urmia University of Technology, Urmia, Iran}\\[0.5em]
$^{6}${\it Department of Physics, Salman Farsi University of Kazerun, Kazerun, Iran}\\[0.5em]
$^{7}${\it Faculty of Informatics, Masaryk University, Botanick\'a 68a, 60200 Brno, Czech Republic}\\[0.5em]

\end{center}
\baselineskip=18pt
\medskip 
\vspace{3cm}
\begin{abstract}
Quantum uncertainty relations are typically analyzed for a pair of incompatible observables, however, the concept per se naturally extends to situations of more than two observables. 
In this work, we obtain tripartite quantum memory-assisted entropic uncertainty relations and show that the lower bounds of these relations have three terms that depend on the complementarity of the observables, the conditional von-Neumann entropies, the Holevo quantities, and the mutual information. The saturation of these inequalities is analyzed.


\vspace{1cm}
\end{abstract}

\vspace{1cm}

\newpage
 \renewcommand{\thefootnote}{*}
\section{Introduction}
The uncertainty principle is undoubtedly one of the most important topics in quantum theory \cite{Heisenberg}. According to this principle, our ability to predict the measurement outcomes of two incompatible observables, which simultaneously are measured on a quantum system, is restricted. This principle can be stated in various forms, e.g., it can be formulated in terms of the Shannon entropy in quantum information theory. The most famous form of the entropic uncertainty relation (EUR) was introduced by Deutsch \cite{Deutsch}. Later, Maassen and Uffink \cite{Uffink} improved Deutsch's relation. They have shown that for two incompatible observables $X$ and $Z$, the following EUR holds
\begin{equation}\label{Maassen and Uffink}
H(X)+H(Z)\geq -\log_2 (c) \equiv q_{MU},
\end{equation}
where $H(P) = -\sum_{k} p_k \log_2 p_k$ is the Shannon entropy of the measured observable $P \in \left\lbrace  X, Z \right\rbrace $, $p_k$ is the probability of the outcome $k$, and $c = \max_{\left\lbrace \mathbb{X},\mathbb{Z}\right\rbrace } \vert\langle x_{i} \vert z_{j}\rangle \vert ^{2}$, where $\mathbb{X}=\lbrace \vert x_{i}\rangle \rbrace$ and $\mathbb{Z}=\lbrace \vert z_{j}\rangle \rbrace$ are the eigenstates of the observables $X$ and $Z$, respectively. \\
Intriguingly, one can generalize the EUR to the case in the presence of quantum memory by means of an interesting game between two players, Alice and Bob. At the beginning of the game, Bob prepares a quantum state $\rho_{AB}$ and sends the part $A$ to Alice and keeps the part $B$ as a quantum memory. In the next step, Alice carries out a measurement on her quantum system $A$ by choosing one of the observables $X$ and $Z$ and announces her choice to Bob. Bob's task is to predict Alice's measurement outcomes. Berta et al. \cite{Berta} shown that the bipartite quantum-memory-assisted entropic uncertainty relation (QMA-EUR) is defined as
\begin{equation}\label{Berta}
H(X \vert B)+H(Z \vert B) \geq q_{MU} +S(A \vert B),
\end{equation}
where $H(P \vert B) = S(\rho_{PB})-S(\rho_{B})$ is the conditional von-Neumann entropy of the post-measurement state after measuring $P$ ($X$ or $Z$) on the part $A$,
\begin{equation}
\rho_{XB}= \sum_{i}(\vert x_{i}\rangle\langle x_{i}\vert_{A}\otimes \mathbf{I}_B ) \rho_{AB}(\vert x_{i}\rangle\langle x_{i}\vert_{A}\otimes \mathbf{I}_{B} ),\nonumber
\end{equation}
\begin{equation}
\rho_{ZB}= \sum_{j}(\vert z_{j}\rangle\langle z_{j}\vert_{A}\otimes \mathbf{I}_{B} ) \rho_{AB}(\vert z_{j}\rangle\langle z_{j}\vert_{A}\otimes \mathbf{I}_{B} ),\nonumber
\end{equation}
and $S(A|B) = S(\rho_{AB})-S(\rho_{B})$ is the conditional von-Neumann entropy with $\rho_{B}=tr_{A}(\rho_{AB})$.
The QMA-EUR has many potential applications in various quantum information processing tasks, such as quantum key distribution \cite{Berta,Koashi}, quantum cryptography \cite{Dupuis,Koenig}, quantum randomness \cite{Vallone,Cao}, entanglement witness \cite{Berta2,Huang}, EPR steering \cite{Walborn,Schneeloch}, and quantum metrology \cite{Giovannetti}.
Due to its importance in quantum information processing, much efforts have been made to expand and improve this relation \cite{Renes,Coles1,Bialynicki,Wehner,Pati,Ballester,Vi,Wu,Rudnicki,Pramanik,Maccone,Coles,Adabi,Adabi1,Dolat2,Dolatkhah,Haseli2,Haseli111,Zozor,R,Kamil,Rudnicki1,Pramanik1}. Impressively, many authors attempted to establish the relationship between quantum correlations and entropic uncertainty relations \cite{r39,r40,r41,r46,r47,r48,r50,r53,r60,r65,r66,r69,r70,n2,n4,rssm,rssm2,rssm3,new02,new04,new05,new06,new07,new01,wangr,Dolat222,hadoqe}.\\
The bipartite QMA-EUR can be extended to the tripartite one in which two additional particles $B$ and $C$ are considered as the quantum memories. Moreover, Haddadi et al. \cite{hadsci} proposed a new QMA-EUR for multipartite systems where the memory is split into multiple parts. In the tripartite scenario, Alice, Bob, and Charlie share a quantum state $\rho_{ABC}$ and Alice performs one of two measurements, $X$ and $Z$, on her system. If Alice measures $X$, then Bob's task is to minimize his uncertainty about $X$. Also, if she measures $Z$, then Charlie's task is to reduce his uncertainty about $Z$. Explicitly, the tripartite QMA-EUR is expressed as \cite{Renes,Berta}
\begin{equation}\label{tpu}
H(X \vert B)+H(Z \vert C)\geq q_{MU}.
\end{equation}
The tripartite QMA-EUR has important applications in quantum information science, such as quantum key distribution \cite{Berta}. However, it is good to know that there have been few improvements to the tripartite QMA-EURs (see, for example, Refs. \cite{Ming,Dolat}). Recently, the lower bound of the tripartite QMA-EUR is improved by adding two additional terms to the lower bound of inequality (\ref{tpu}), viz \cite{Dolat}
\begin{equation}\label{tpu3}
H(X \vert B)+H(Z \vert C)\geqslant q_{MU}+\frac{S(A|B)+S(A|C)}{2}+\max\{0 , \delta\},
\end{equation}
with $$\delta=\frac{I(A:B)+I(A:C)}{2} -[I(X:B)+I(Z:C)],$$
where $$I(A:B(C))=S(\rho_{A})+S(\rho_{B(C)})-S(\rho_{AB(C)}),$$ is mutual information between Alice and Bob (Charlie), furthermore
$$I(P:B(C))= S(\rho_{B(C)})- \sum_{i}p_{i}S(\rho_{B(C)|i})$$ is the Holevo quantity which is equal to the upper bound of the accessible information to Bob (Charlie) about Alice's measurement outcomes.\\
\\
In recent studies \cite{Dolat,sr2021,hadwang}, it is shown that this lower bound is tighter than the other bounds that have been introduced so far.\\
Up to now, we have considered only EURs and QMA-EURs with two measurements (observables). However, one can generalize QMA-EURs to more than two measurements. Recently, the QMA-EURs for multiple measurements have attracted increasing interest. Many bipartite QMA-EURs for more than two observables have been obtained \cite{Dolatkhah,Liu,Zhang,Yunlong,multi2,multi3,multi4,multi5,multi6,multi7}. Nevertheless, according to our knowledge so far, no relation has been obtained for the tripartite QMA-EUR with multiple measurements.
Motivated by this, we obtain several tripartite QMA-EURs for multiple measurements. The lower bounds of these relations have three terms that depend on the observables' complementarity, the conditional von-Neumann entropies, the Holevo quantities, and the mutual information. It is hoped that these relations have many potential wide applications in quantum theory, and expect that these relations can be demonstrated in many physical systems.\\
The paper is organized as follows: In Sec. \ref{sec2}, several lower bounds are introduced for the tripartite QMA-EUR with multiple measurements. In Sec. \ref{sec3}, these lower bounds are examined through two cases. Lastly, the results are summarized in Sec. \ref{sec4}.

\section{Tripartite QMA-EURs for multiple measurements}\label{sec2}
In this section, several tripartite QMA-EURs for multiple measurements are derived by utilizing the relevant bounds for the sum of Shannon entropies. As mentioned earlier, several EURs for multiple measurements have been proposed \cite{Liu,Zhang,Yunlong,multi2,multi3,multi4,multi5,multi6,multi7}. For example, Liu et al. \cite{Liu} obtained an EUR for $N$ measurements $(M_{m}, m=1,2,...,N)$ as follows
\begin{equation}\label{Liu}
\sum^{N}_{m=1} H(M_{m})\geq -\log_{2}(b)+(N-1)S(\rho_{A}),
\end{equation}
where $ S(\rho_{A})=-tr(\rho_{A}\log_2 \rho_{A}) $ is the von-Neumann entropy of measured system $\rho$ and
$$b=\max_{i_{N}} \left\{ \sum_{i_{2}\sim{i_{N-1}}} \max_{i_{1}} \Big[ |\langle u^{1}_{i_{1}}|u^{2}_{i_{2}}\rangle |^{2} \Big]\prod^{N-1}_{m=2}|\langle u^{m}_{i_{m}}|u^{m+1}_{i_{m+1}}\rangle |^{2} \right\} ,$$
in which $|u^{m}_{i_{m}}\rangle$ is the $i$th eigenvector of $M_{m}$. Besides, the EUR for $N$ measurements obtained by Zhang et al. \cite{Zhang} is
\begin{equation}\label{Zhang}
\sum^{N}_{m=1} H(M_{m})\geq(N-1)S(\rho_{A})+\max_{u}\lbrace \ell^{U}_{u}\rbrace ,
\end{equation}
where
$$\ell^{U}_{u}=-\sum_{i_{N}}p_{u^{{N}}_{i_{N}}}\log_{2} \sum_{i_{k},N\geq k>1} \max_{i_{1}} \prod^{N-1}_{m=1}|\langle u^{m}_{i_{m}}|u^{m+1}_{i_{m+1}}\rangle|^{2},$$
and
$$p_{u^{{N}}_{i_{N}}}=tr \big[ |u^{N}_{i_{N}}\rangle\langle u^{N}_{i_{N}}| \rho_{A} \big].$$
In another case, Xiao et al. \cite{Yunlong} derived the following EUR for multiple measurements as
\begin{equation}\label{Xiao}
\sum^{N}_{m=1} H(M_{m})\geq(N-1)S(\rho_{A})-\frac{1}{N}\omega \mathfrak{B},
\end{equation}
where $\mathfrak{B}$ denotes the certain vector of logarithmic distributions and $\omega$ shows the universal majorization bound of $N$ measurements.

Now, we show that it is possible to use the lower bounds of the above-mentioned relations to obtain tripartite QMA-EURs for multiple measurements.

\begin{figure}[H]
  \centering
  \includegraphics[width=0.70\textwidth]{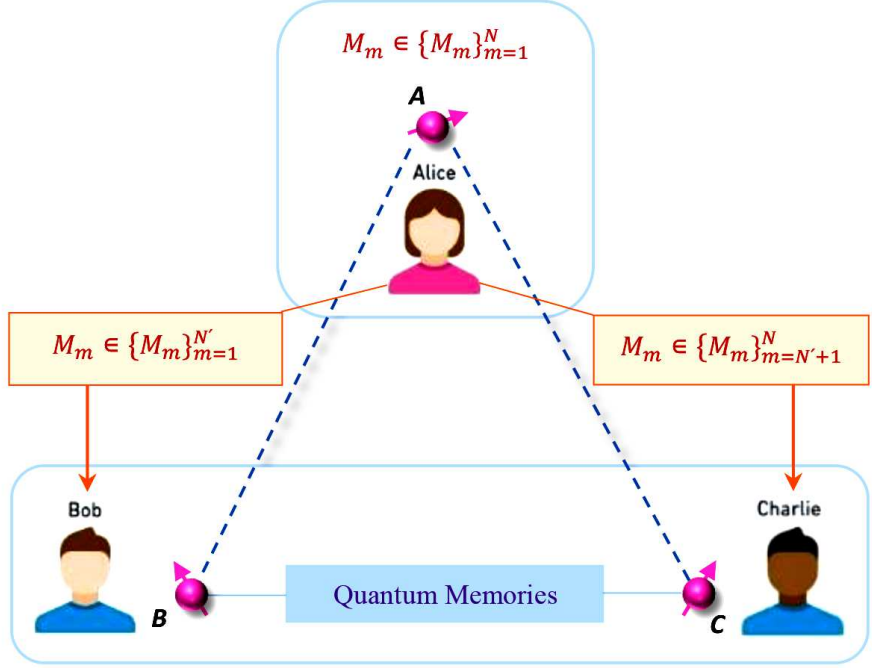}
\caption{A schematic representation of the tripartite uncertainty game for multiple measurements. The dashed lines denote the quantum correlation between the particles.}
\label{fig:1}
\end{figure}

\noindent\textbf{Definition}  (The tripartite guessing game with multiple measurements).  Let us consider a tripartite uncertainty game (so-called monogamy game) between Alice, Bob, and Charlie. Before the game, Alice, Bob, and Charlie agree on a set of measurements $(\lbrace M_{m}\rbrace , m=1,2,...,N)$. Then, they share a tripartite quantum state $\rho_{ABC}$. Alice does her measurement on her quantum system ($A$) with one of the measurements. If Alice measures one of $N^{'} (N^{'}<N)$ measurements $(\lbrace M_{m}\rbrace , m=1,2,...,N^{'})$, then Bob's task is to minimize his uncertainty about Alice’s measurement outcomes. And if she measures one of $N-N^{'}$ measurements $(\lbrace M_{m}\rbrace , m=N^{'}+1,...,N)$, then Charlie’s task is to minimize his uncertainty about Alice’s measurement outcomes. This scenario is displayed in Fig. \ref{fig:1}.\\

\noindent\textbf{Theorem 1.} The following tripartite QMA-EUR with multiple measurements holds for any quantum state $\rho_{ABC}$
\begin{equation}\label{Theorem1}
\sum^{N^{\prime}}_{m=1}H(M_{m}\vert B)+\sum^{N}_{m=N^{\prime}+1}H(M_{m}\vert C)\geq -\log_{2}(b)+(N-1)\frac{S(A\vert B)+S(A\vert C)}{2}+\max\{0 , \delta\},
\end{equation}
where
\begin{equation}\label{delta1}
\delta=(N-1)\frac{I(A:B)+I(A:C)}{2}-\lbrace\sum^{N^{\prime}}_{m=1}I(M_{m}:B)+\sum^{N}_{m=N^{\prime}+1}I(M_{m}:C)\rbrace.
\end{equation}

\vspace{0.5cm}
\noindent\textbf{Proof.} Regarding inequality (\ref{Liu}), one can obtain a tripartite QMA-EUR for multiple measurements. To achieve this aim, one can use the definition of the von-Neumann conditional entropy $H(M_{m} \vert B (C)) = S(\rho_{M_{m}B(C)})-S(\rho_{B(C)})$ and that of the mutual information $I(M_{m}:B(C))=H(M_{m})+S(\rho_{B(C)})-S(\rho_{M_{m}B(C)})$. Adding the two quantities for $N$ measurements, one obtains
\begin{equation}\label{Liu666}
\sum^{N}_{m=1} H(M_{m})=\sum^{N^{\prime}}_{m=1}H(M_{m}\vert B)+\sum^{N}_{m=N^{\prime}+1}H(M_{m}\vert C)+\sum^{N^{\prime}}_{m=1}I(M_{m}:B)+\sum^{N}_{m=N^{\prime}+1}I(M_{m}:C).
\end{equation}
By substituting Eq. (\ref{Liu666}) into Eq. (\ref{Liu}), one obtains
\begin{eqnarray}\label{Liu 33}
\sum^{N^{\prime}}_{m=1}H(M_{m}\vert B)+\sum^{N}_{m=N^{\prime}+1}H(M_{m}\vert C)&\geq &-\log_{2}(b)+(N-1)S(\rho_{A})\nonumber\\ & & -\sum^{N^{\prime}}_{m=1}I(M_{m}:B)-\sum^{N}_{m=N^{\prime}+1}I(M_{m}:C).
\end{eqnarray}
Using \cite{Nielsen}
\begin{equation}\label{aaaa}
S(\rho_{A})=\frac{S(A\vert B)+S(A\vert C)}{2} +\frac{I(A:B)+I(A:C)}{2},
\end{equation}
in Eq. (\ref{Liu 33}), one comes to
\begin{eqnarray}\label{Liu 3}
\sum^{N^{\prime}}_{m=1}H(M_{m}\vert B)+\sum^{N}_{m=N^{\prime}+1}H(M_{m}\vert C)  \geq & -&\log_{2}(b)+(N-1)\frac{S(A\vert B)+S(A\vert C)}{2} \nonumber\\
 &+&(N-1)\frac{I(A:B)+I(A:C)}{2} \nonumber\\ &-&\sum^{N^{\prime}}_{m=1}I(M_{m}:B)-\sum^{N}_{m=N^{\prime}+1}I(M_{m}:C),
\end{eqnarray}
which can be formulated as the desired result (\ref{Theorem1}).\\
It is worth noting that the second term of this lower bound (\ref{Theorem1}) is always non-negative due to the strong subadditivity inequality \cite{Nielsen}. In other words, the second term of this lower bound states that strong subadditivity inequality plays role in tripartite QMA-EUR for multiple measurements.\\
\\
\textbf{Corollary 1.} One finds another tripartite QMA-EUR for multiple measurements as
\begin{equation}\label{zhang 2}
\sum^{N^{\prime}}_{m=1}H(M_{m}\vert B)+\sum^{N}_{m=N^{\prime}+1}H(M_{m}\vert C)\geq \max_{u}\lbrace \ell^{U}_{u} \rbrace +(N-1)\frac{S(A\vert B)+S(A\vert C)}{2} +\max\{0 , \delta\},
\end{equation}
where $\delta$ is the same as that in Eq. (\ref{delta1}).\\
\\
\textbf{Proof.}  With the help of Eqs. (\ref{Zhang}), (\ref{Liu666}), and (\ref{aaaa}), we obtain the inequality (\ref{zhang 2}).\\
\\
\textbf{Corollary 2.} One obtains a new tripartite QMA-EUR for multiple measurements as
\begin{equation}\label{Xiao 2}
\sum^{N^{\prime}}_{m=1}H(M_{m}\vert B)+\sum^{N}_{m=N^{\prime}+1}H(M_{m}\vert C)\geq -\frac{1}{N}\omega \mathfrak{B}+(N-1)\frac{S(A\vert B)+S(A\vert C)}{2}+\max\{0 , \delta\},
\end{equation}
where $\delta$ is similar to that in Eq. (\ref{delta1}).\\
\\
\textbf{Proof.} In the same way, by employing Eqs. (\ref{Xiao}), (\ref{Liu666}), and (\ref{aaaa}), it is easy to restore the inequality (\ref{Xiao 2}).\\
\\
\textbf{Generalization} (Tripartite QMA-EUR for all entropy-based uncertainty relations with multiple measurements). Based on what has been mentioned so far, it is obvious that any entropic uncertainty relation for multiple measurements can be easily converted to tripartite QMA-EUR for multiple measurements. To do so, assume that the general form of the uncertainty relation for $N$ measurements $(M_{1},M_{2},$ ...,$M_{N})$ is
\begin{equation}\label{general}
\sum^{N}_{m=1} H(M_{m})\geq LB,
\end{equation}
where $LB$ is an abbreviation for lower bound. Equipped with Eq. (\ref{Liu666}), this relation can be transformed into
\begin{equation}\label{general 2}
\sum^{N^{\prime}}_{m=1}H(M_{m}\vert B)+\sum^{N}_{m=N^{\prime}+1}H(M_{m}\vert C)\geq LB-\sum^{N^{\prime}}_{m=1}I(M_{m}:B)-\sum^{N}_{m=N^{\prime}+1}I(M_{m}:C),
\end{equation}
which is a tripartite QMA-EUR for multiple measurements. As can be seen, it is a simple way which can be used to convert an entropy-based uncertainty relation in the absence of quantum memory to the tripartite QMA-EUR. For example, Coles et al. \cite{Coles44} derived the following EUR for any state $\rho$ of a qubit and any complete set of three mutually unbiased observables $x$, $y$, and $z$, i.e.
\begin{equation}
H(x)+H(y)+H(z)\geq 2 \log_{2} 2 +S(\rho_{A}).
\end{equation}
By using Eqs. (\ref{aaaa}) and (\ref{general 2}), this relation can be converted into a tripartite QMA-EUR for any state $\rho_{ABC}$ of a three-qubit and any complete set of three mutually unbiased observables $x$, $y$, and $z$ as
\begin{equation}\label{coles}
H(x\vert B)+H(y\vert C)+H(z\vert C)\geq   \log_{2} 2 +\frac{S(A\vert B)+S(A\vert C)}{2} +\max\{0 , \delta^{\prime}\},
\end{equation}
where $$\delta^{\prime}=\log_{2} 2 + \frac{I(A:B)+I(A:C)}{2}-\lbrace I(x:B)+I(y:C)+I(z:C)\rbrace.$$
It is interesting to note that the lower bound of Eq. (\ref{coles}) is perfectly tight for the class of three-qubit states with maximally mixed subsystem $A$. Also, when subsystem $A$ is diagonal in the eigenbasis one of the observables  $x$, $y$, and $z$, one can conclude that the lower bound is again perfectly tight.

\section{Examples}\label{sec3}
For simplicity, let us consider three mutually unbiased observables $x=\sigma_x$, $y=\sigma_y$, and $z=\sigma_z$ measured on the part $A$ of a three-qubit state $\rho_{ABC}$. Herein, we consider two different cases.\\

\vspace{0.3cm}

\noindent\textbf{Case 1.}  It is assumed that if Alice measures observable $\sigma_x$ or $\sigma_y$, then Bob's task is to guess the results of Alice's measurement. And if she measures observable $\sigma_z$, then Charlie's task is to guess Alice's measurement results. So, the tripartite uncertainty ($U$) and its lower bounds ($L_{1}$ and $L_{2}$) based on Eqs. (\ref{Theorem1}) and  (\ref{coles}) for this case are
\begin{equation}\label{U1}
U=H(\sigma_x\vert B)+H(\sigma_y\vert B)+H(\sigma_z\vert C),
\end{equation}

\begin{equation}\label{L1}
 L_{1}=\log_2 2 +S(A\vert B)+S(A\vert C) +\max\{0 , \kappa\},
\end{equation}
and
\begin{equation}\label{L2}
L_{2}=\log_2 2 +\frac{S(A\vert B)+S(A\vert C)}{2} +\max\{0 , \kappa^{\prime}\},
\end{equation}
respectively, where $$\kappa=I(A:B)+I(A:C)-\lbrace I(\sigma_x:B)+I(\sigma_y:B)+I(\sigma_z:C)\rbrace,$$ and $$\kappa^{\prime}=\log_2 2+\frac{I(A:B)+I(A:C)}{2}-\lbrace I(\sigma_x:B)+I(\sigma_y:B)+I(\sigma_z:C)\rbrace.$$

\noindent\textbf{Case 2.}  It is supposed that if Alice measures observable $\sigma_x$, then Bob's task is to guess the results of Alice's measurement, and if she measures observable $\sigma_y$ or $\sigma_z$, then Charlie's task is to guess Alice's measurement results. Then, tripartite uncertainty and its lower bounds according to Eqs. (\ref{Theorem1}) and  (\ref{coles}) for this case would be
\begin{equation}\label{Up1}
U^{\prime}=H(\sigma_x\vert B)+H(\sigma_y\vert C)+H(\sigma_z\vert C),
\end{equation}

\begin{equation}\label{L1p}
L^{\prime}_{1} = \log_2 2 +S(A\vert B)+S(A\vert C) +\max\{0 , \eta\},
\end{equation}
and
\begin{equation}\label{L2p}
L^{\prime}_{2}=\log_2 2 +\frac{S(A\vert B)+S(A\vert C)}{2} +\max\{0 , \eta^{\prime}\},
\end{equation}
respectively,
with $$\eta=I(A:B)+I(A:C)-\lbrace I(\sigma_x:B)+I(\sigma_y:C)+I(\sigma_z:C)\rbrace,$$ and $$\eta^{\prime}=\log_2 2+\frac{I(A:B)+I(A:C)}{2}-\lbrace I(\sigma_x:B)+I(\sigma_y:C)+I(\sigma_z:C)\rbrace.$$
Now, let us examine the above cases for two different states that are shared between Alice, Bob, and Charlie.

\subsection{Werner-type state}
As a first example, we consider three observables $x=\sigma_x, y=\sigma_y,$ and $ z=\sigma_z$ measured on the part $A$ of the Werner-type state defined as
\begin{equation}\label{werner}
\rho_{w}=(1-p) \vert GHZ \rangle \langle GHZ \vert + \frac{p}{8}\mathbf{I}_{ABC},
\end{equation}
where $\vert GHZ \rangle = (\vert 000 \rangle + \vert 111 \rangle)/\sqrt{2}$ is the Greenberger-Horne-Zeilinger (GHZ) state and $0 \leq p \leq 1$.

\begin{figure}[H]
\centering
\includegraphics[width=8cm]{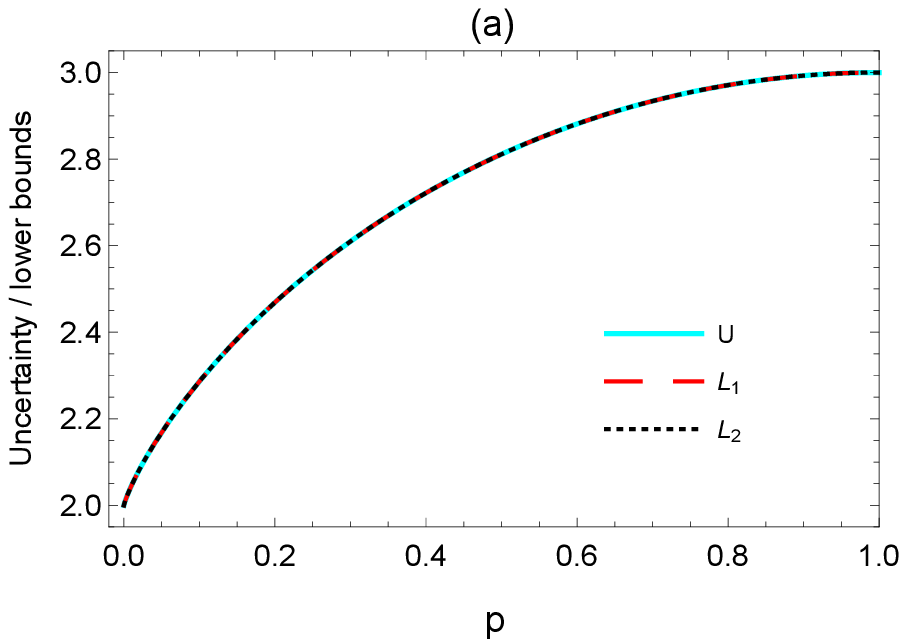}
\hspace{5mm}
\includegraphics[width=8cm]{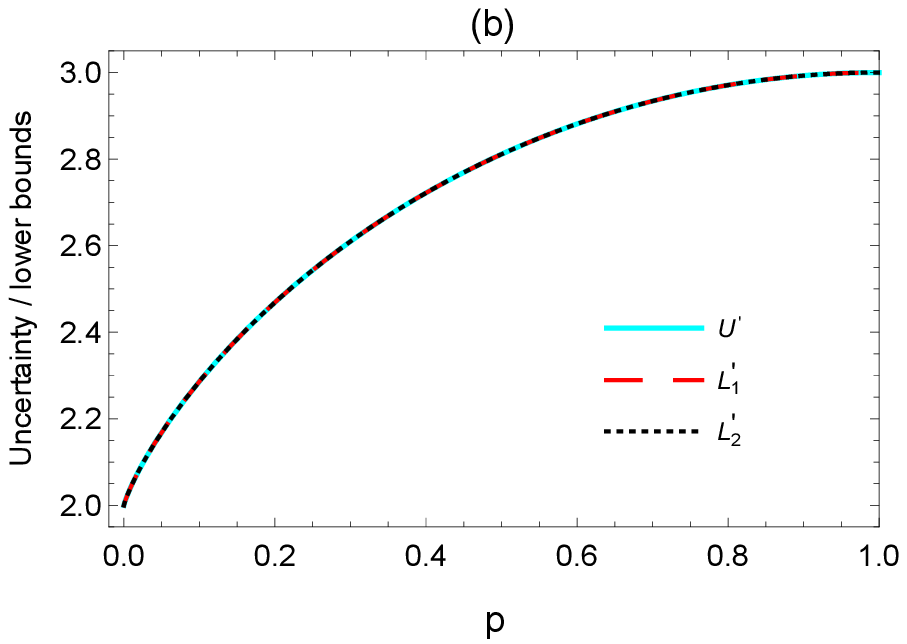}
\caption{Tripartite uncertainty and lower bounds for three complementary observables $\sigma_{x}$, $\sigma_{y}$, and $\sigma_{z}$ measured on the part $A$ of the Werner-type state (Eq. (\ref{werner})) versus the parameter $p$.}\label{fig2}
\end{figure}

In Fig. \ref{fig2}, the tripartite uncertainty and its lower bounds for the measurement of three complementary observables $\sigma_{x}$, $\sigma_{y}$, and $\sigma_{z}$ on the Werner state are plotted versus the parameter $p$. Fig. \ref{fig2}(a) shows the uncertainty ($U$) and lower bounds $(L_{1}$ and $L_{2})$ for case 1. It can be observed that there is a complete overlap between them, $U=L_{1}=L_{2}$ holds. In Fig. \ref{fig2}(b), the measurement uncertainty ($U^{\prime}$) and its lower bounds $(L^{\prime}_{1}$ and $L^{\prime}_{2})$ are considered in the second case. As can be seen, the results are similar to the first case, $U^{\prime}=L^{\prime}_{1}=L^{\prime}_{2}$ is held. From Figs. \ref{fig2}(a) and \ref{fig2}(b), one comes to the result that for Werner-type state, the obtained uncertainty and lower bounds are exactly the same as with each other, that is
\begin{equation}
U=U^{\prime}=L_{1}=L_{1}^{\prime}=L_{2}=L^{\prime}_{2}=-\frac{p}{2}\log_{2}\frac{p}{8}-\frac{2-p}{2}\log_{2}\frac{2-p}{8}.
\end{equation}

\subsection{Generalized $W$ state}
As an another example, we consider the generalized $W$ state defined as
\begin{equation}\label{W}
\vert W_{G} \rangle = \sin\theta \cos\phi \vert 100\rangle +\sin\theta \sin\phi \vert 010\rangle +\cos\theta \vert 001\rangle ,
\end{equation}
where $\theta\in\left[ 0,\pi\right]$ and $\phi\in\left[ 0,2\pi\right)$. The measurement uncertainty and lower bounds of the tripartite QMA-EUR for the above- mentioned two cases are
\begin{equation}
L_{1} = -\log_2 2 +\alpha(\theta)+\beta(\theta)+\gamma_{+}(\theta)+\gamma_{-}(\theta),
\end{equation}
\begin{equation}
U=L_{2} = \alpha^{\prime}(\theta)+2\beta(\theta)+\gamma_{+}(\theta)+\gamma_{-}(\theta),
\end{equation}
\begin{equation}
L^{\prime}_{1} = \alpha^{\prime}(\theta)+\zeta(\theta)+ \gamma_{+}(\theta)+ \gamma_{-}(\theta),
\end{equation}
\begin{equation}
U^{\prime}=L^{\prime}_{2} = \alpha^{\prime \prime}(\theta)+\zeta^{\prime}(\theta)+ \beta(\theta)+\gamma_{+}(\theta)+ \gamma_{-}(\theta),
\end{equation}

where $\phi=\pi/4$ and
$$
\begin{aligned}
&\alpha(\theta)=\frac{\sin^{2}(\theta)}{2}\log_2[2 \sin^{2}(\theta)], \quad \beta(\theta)=\frac{2\cos^{2}(\theta)+\sin^{2}(\theta)}{2}\log_{2}[\frac{2\cos^{2}(\theta)+\sin^{2}(\theta)}{2}],\\&
\gamma_{\pm}(\theta)=-\frac{2\pm\sqrt{3+\cos(4\theta)}}{2}\log_{2}[\frac{2\pm\sqrt{3+\cos(4\theta)}}{8}],\\&
\alpha^{\prime}(\theta)=\sin^{2}(\theta)\log_2[\sin^{2}(\theta)],\quad \zeta=\cos^{2}(\theta)\log_{2}[\frac{\cos^{2}(\theta)}{2}],\\&
\alpha^{\prime \prime}(\theta)=\frac{\sin^{2}(\theta)}{2}\log_2[2 \sin^{6}(\theta)], \quad    \zeta^{\prime}(\theta)=\cos^{2}(\theta)\log_{2}[\cos^{2}(\theta)].\end{aligned}
$$

\begin{figure}[H]
\centering
\includegraphics[width=8cm]{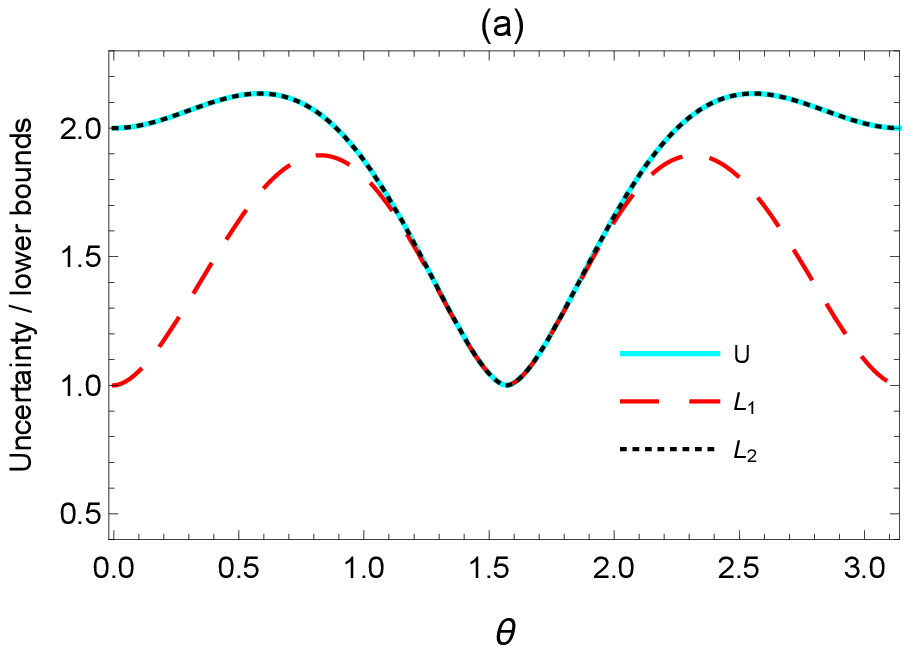}
\hspace{5mm}
\includegraphics[width=8cm]{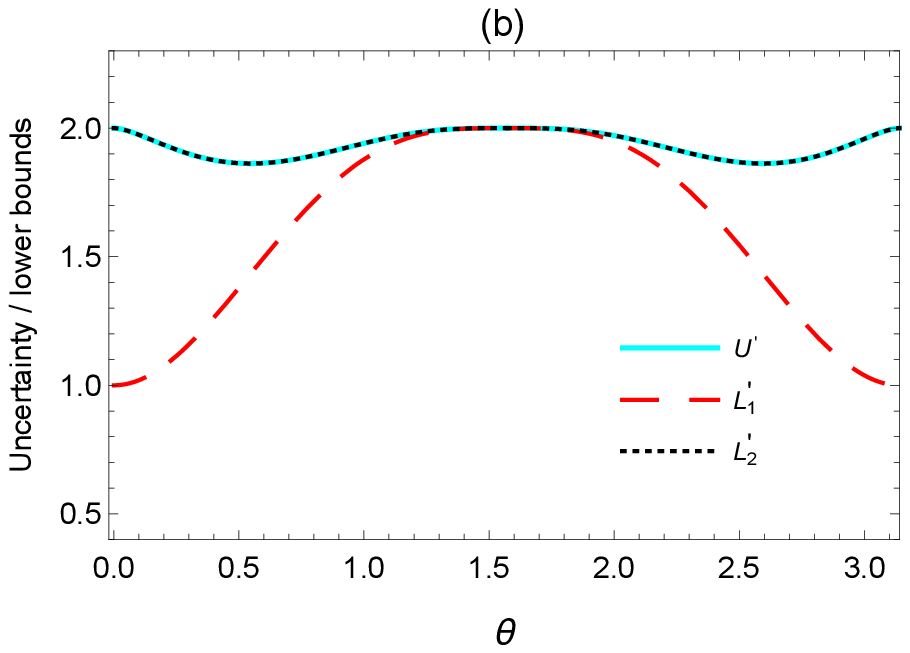}
\caption{Tripartite uncertainty and lower bounds for three complementary observables $\sigma_{x}$, $\sigma_{y}$, and $\sigma_{z}$ measured on the part $A$ of the generalized $W$ state (Eq. (\ref{W})) versus the parameter $\theta$ with $\phi=\pi/4$.}\label{fig3}
\end{figure}

In Fig. \ref{fig3}, the uncertainty and lower bounds of the tripartite QMA-EUR for the measurement of three complementary observables $\sigma_{x}$, $\sigma_{y}$, and $\sigma_{z}$ on this state are plotted versus the parameter $\theta$.  By comparing Figs. \ref{fig3}(a) and \ref{fig3}(b), it is clear that in contrast to the first example, the lower bounds for the above-mentioned two cases are not the same. This shows that the lower bounds not only depend on the initial state but also on the observables.

After a short algebra it follows from Eqs. (\ref{L1}, \ref{L2}) and (\ref{L1p}, \ref{L2p}) that
\begin{equation}
 L_{2}= L_{1}+\log_2 2-S(\rho_{A}), \quad \textmd{and} \quad L_{2}^{\prime}= L_{1}^{\prime}+\log_2 2-S(\rho_{A})\,.
\end{equation}
Thus, for all states shared by Alice, Bob and Charlie, $L_{2}(L^{\prime}_{2})$ is tighter than $L_{1}(L^{\prime}_{1})$, because $\log_2 2-S(\rho_{A})\geq 0$. It should be mentioned that when the subsystem $A$ is in a maximally mixed state, these lower bounds coincide. The differences between these lower bounds increase with the decreasing purity of $\varrho_A$ and achieve their maximum for pure states. This monotonicity can be seen in all figures.

These examples show that considering the importance of the initial state shared between Alice, Bob, and Charlie, the order of the chosen three complementary observables does also affect changing the result. Therefore, the best measurement precision can be obtained by choosing the proper initial state and complementary observables.

\section{Conclusion}\label{sec4}
In summary, we have presented several tripartite QMA-EURs for multiple measurements settings. The terms of lower bounds depend on the complementarity of the observables, the conditional von-Neumann entropies, the Holevo quantities and the mutual information. It should be mentioned that one of the terms in obtained lower bounds is closely related to strong subadditivity inequality. As illustrations, the lower bounds have been examined for the Werner-type and generalized $W$ states.  Accordingly, the results have shown that the selection of the appropriate initial state, as well as the complementary observables, have a notable effect on the measurement uncertainty. Our new lower bounds are expected to be useful in various quantum information processing tasks and may also provide a better perception of the lower bounds of the tripartite QMA-EUR which are important in improving quantum measurements precision. For example, our lower bounds can more accurately capture the tradeoff of entanglement monogamy which increases the prediction precision of measurement results \cite{Coles1}. Furthermore, we think that the experimental realization of our lower bounds can be implemented for more than two measurements settings in the presence of quantum memories.

\vspace{20pt}

\section*{ORCID iDs}
Hazhir Dolatkhah \href{https://orcid.org/0000-0002-2411-8690}{https://orcid.org/0000-0002-2411-8690}\\
Saeed Haddadi \href{https://orcid.org/0000-0002-1596-0763}{https://orcid.org/0000-0002-1596-0763}\\
Soroush Haseli \href{https://orcid.org/0000-0003-1031-4815}{https://orcid.org/0000-0003-1031-4815}\\
Mohammad Reza Pourkarimi \href{https://orcid.org/0000-0002-8554-1396}{https://orcid.org/0000-0002-8554-1396}\\
Mario Ziman \href{https://orcid.org/0000-0002-4805-5890}{https://orcid.org/0000-0002-4805-5890}\\

\section*{Disclosures}
The authors declare no conflicts of interest.

\section*{Acknowledgements} The support of projects APVV-18-0518 (OPTIQUTE) and VEGA 2/0161/19 (HOQIT) is acknowledged.

\end{document}